\documentclass[12pt,preprint]{aastex}
\usepackage{color}

\begin{document}

\title{Observations of Multiple Surges Associated with Magnetic Activities in AR10484 on 25 October 2003}

\author{Wahab Uddin\altaffilmark{1}, B.~Schmieder\altaffilmark{2}, R. Chandra\altaffilmark{3}, Abhishek K.~Srivastava\altaffilmark{1}, Pankaj~Kumar\altaffilmark{4}, S. Bisht\altaffilmark{3}}

\altaffiltext{1}{Aryabhatta Research Institute of Observational Sciences (ARIES), Nainital, India.}
\altaffiltext{2}{LESIA, Observatoire de Paris-Meudon, 92195, Meudon Cedex, France.}
\altaffiltext{3}{Department of Physics, DSB Campus, Kumaun University, Nainital, India, 263 002.}
\altaffiltext{4}{Korea Astronomy and Space Science Institute (KASI), Daejeon, 305-348, Republic of Korea.}
\email{wahab@aries.res.in}

%
\begin{abstract}

We present a multiwavelength study of recurrent surges
observed in H$\alpha$, UV (SOHO/EIT) and Radio (Learmonth, Australia)
from the super-active region NOAA 10484 on 25 October,
2003.
Several bright structures visible  in H$\alpha$ and UV corresponding
to subflares are also observed at the base of each surge. Type III
bursts are triggered and RHESSI X-ray sources are evident with
surge activity. The major surge  consists of  the bunches of
ejective paths forming a fan-shape region with an angular size of
($\approx 65^\circ$) during its maximum phase.  
The ejection speed reaches upto $\sim$200 km/s.
The SOHO/MDI magnetograms reveal that a large  dipole emerges east side of the active
region on 18-20 October 2003, a few days before the surges. On  October 25, 2003, 
the major sunspots were  surrounded by $"$moat regions$"$
with moving magnetic features (MMFs).
Parasitic fragmented positive polarities were pushed by the ambient
dispersion motion of the MMFs and annihilated with  
negative polarities  at the borders of the moat region of the following spot
to produce 
flares and surges. 
A topology analysis of the global Sun using PFSS shows that the fan structures
visible in the EIT 171 \AA\ images follow magnetic field lines connecting the
present AR to a preceding AR in the South East. Radio observations of type III
bursts indicate that they are coincident with the surges, suggesting that
magnetic reconnection is the driver mechanism. The magnetic energy released
by reconnection is transformed into plasma heating and provides the kinetic
energy for the ejections. A lack of a radio signature in the high corona
suggests that the surges are confined to follow the closed field lines in the
fans. We conclude that these cool surges may have some local heating effects in
the closed loops, but probably play a minor role in global coronal heating and
the surge material does not escape to the solar wind.

\end{abstract}
\section{INTRODUCTION}
Solar surge is a collimated ejection of plasma material from the
lower solar atmosphere into the corona. These ejecta exhibit
episodic heating and cooling also, therefore, may be visible in the
range of emissions from H${\alpha}$ to EUV/UV, and X-rays, 
and can be abbreviated in general as solar jets \citep{schmieder95}. The surges may have
initiation velocity of $\sim$50 km s$^{-1}$, which may further
increase up to a maximum value of 100$-$300 km s$^{-1}$, and
these surges may 
reach up to the heights of 10 Mm to 200 Mm or even more
\citep{Ster20}. Lifetime of surges is about 30 min, and 
they can be recurrent
with a period of an hour or more \citep{schmieder84,schmieder95}.
Usually the surge is confined to one or several narrow threads of
magnetic fields embedded in the plasma that shoot out 
above the solar
surface. However, such surges are mostly associated with the flaring
regions and the sites of solar transients where recurrent magnetic
reconnection 
is dominant. The evolution of solar surges 
has been studied comprehensively 
in association with magnetic field emergence and cancellation, as well
as flaring activities of the solar
atmosphere where such plasma jets also appeared twisted and spiraled
\citep[e.g.,][and references cited there in]
{Sch94,Chae99,Yushi03,Liu04}.

The solar surges may occur in the regions of emerging magnetic
fluxes in the vicinity of satellite spots
\citep{rust1968,roy1973,kurokawa93}. Often these surges are
associated with flares \citep{schmieder88,schmieder95,uddin04, chandra06}, and
magnetic reconnection may be responsible for the acceleration as
well as heating of the plasma. 
An other kind of reconnection could occur due to the collision of opposite polarity magnetic fluxes
in $"$moat region$"$ \citep{Brooks07}. Small moving magnetic features called
MMFs \citep{Harvey73,Dalda07,Kitiashvili10}, are observed  
as moat regions. The formation of MMFs is closely related to the
fragmentation and disintegration of sunspot magnetic fluxes. Recent
studies show that the amount of magnetic flux 
lost by the sunspots
is similar to  the flux transported in the moat region
\citep{Kubo08}. The flux is annihilated at the border of the moat
region, and in its consequence the subflares, jets and surges may occur
during the reconnection processes along neutral lines  \citep{Beck07, Brooks07, brooks08, engell11}.
Theoretical models have been developed concerning canceling flux producing surges or jets.
The emerging-flux model of \citet{yokoyama1996} supports this kind of surge dynamics, which may
be triggered due to the interaction of emerging photospheric field
with the pre-existing overlying coronal magnetic fields. Although
magnetic reconnection and photospheric magnetic activities may be
the key in driving many solar surges and other jets, however,
several other mechanisms may also be responsible for the surge/jet
dynamics. \cite{pariat10} have developed  a 3D reconnection
model without evidence of emerging flux. This model is able to
generate untwisting jets when a stress is constantly applied at the
photosphere \citep{rachmeler10}. \cite{Shib82} and \cite{Ster93}
have reported that the pressure pulse can trigger the solar surges
of moderate heights in the hydrodynamic regime of the 
solar atmosphere. Solar surges may also be accelerated due to the
whip motion of the reconnection generated newly formed 
magnetic field \citep{Shib92, Can96}, while the reconnection
generated explosive events may also trigger such kind of plasma
dynamics \citep{Maj09}. In addition to the typical solar surges,
\cite{Gu99} have observed the polar surges as cool
jets at polar region without any association with transients. 
Recently, the cool jets and surges have also been modeled respectively
in the polar region as well as near the boundary of a non-flaring
active region due to
reconnection generated velocity pulses in the ideal
magnetohydrodynamic (MHD) regime of the solar atmosphere \citep{Sri11,Pra12}. In
conclusions, the solar surges and other various types of solar jets
may be excited via both, e.g., the direct magnetic reconnection
processes in the emerging field regions, as well as due to the
magnetohydrodynamic wave activities.

During October-November 2003, major solar activity originated from
three super-active regions, namely NOAA AR 10484, 10486 and 10488.
The active region NOAA 10484 (N05W29) evolved on 25 October, 2003 was very
complex having $\beta\gamma\delta$ configuration. This AR has
produced many recurrent surges and flare activities during its
passage on the solar disc. It produced major surge activities on 22
and 25 October, 2003. On October 25, we observed recurrent surges
between 01:50 UT and 04:15 UT. A preliminary report on these
observations has been presented in \citet{Uddin10}. In this active
region, there was no evidence of strong emerging magnetic fluxes
during the recurrent surges.  

Many questions arise around the surge activity:

 What is the trigger mechanism of the surges?
Are they due to reconnection with the pre-existing field lines?
What is the dynamics of the magnetic boundary  bringing   collision of opposite polarities?
Are the pre-existing field lines open or close at the periphery of the active region? 
Commonly outflows are observed at the periphery of active regions  \citep{harra08, delzanna08}.
\citet{mcintosh09} claimed that dynamic chromospheric  spicules in the outskirts of ARs are related to these outflows.
\citet{warren11} and \citet{ugarte11} claimed that there is no direct relationship between these two populations of structures,
i.e., the hot and cool loops maintained respectively at mega/sub-mega-Kelvin temperatures.
The questions arises that whether these observed 
surges in AR 10484 may 
participate to the commonly  observed outflows in the outskirts of active regions or not?
To understand this issue, statistical studies should be done to find  if any relationship  between these cool jets and  the
outflows of hot plasma commonly observed at the periphery of active regions does exist.
However, this topic is out of the scope of present work.

In this paper, we present a detail multiwavelength study of recurrent
surges and their associated events (e.g., flares) as occurred in AR 10484. In
Section~\ref{obs}, we present the details of data sets used in this
study. The multiwavelength evolution of the surges and their
association with subflares are described in Section~\ref{multi}. In
section~\ref{mag}, we describe the magnetic field evolution of
active region before and during the surges and associated flares. We discuss  on the
possibility for a decaying active region and its magnetic activities
to produce surges,  and conclude, in the last section, on our results in the frame of the above questions
 and on  the possible trigger of the surges: wave or reconnection .



\section{OBSERVATIONAL DATA SETS}
\label{obs}
The data sets used for our present study, have been taken from following sources:\\

\noindent $\bf {H\alpha~ Data:}$  The H$\alpha$ observations of the flare and associated surges
were carried out at ARIES, Nainital, India by using 15 cm f/15 Coud\'e
solar tower telescope equipped with H$\alpha$ filter. The image size was
enlarged by a factor of two using a Barlow lens. The images were
recorded by a 16 bit 576$\times$384 pixels CCD camera system having
pixel size 22 $\rm{\mu^2}$. The resolution of the images
is 1$^{\prime\prime}$ per pixel. The cadence for the images is $\sim$15-20 sec.

\noindent $\bf {SOHO/MDI~ Data:}$
To understand the evolution of magnetic complexity of the active region, we use SOHO/MDI data. The magnetic field
data was taken from the SOHO/MDI instrument \citep{sche1995}. The cadence of images is 96~minute and the pixel resolution
is 1.98$^{\prime\prime}$.

\noindent $\bf {SOHO/EIT~ Data:}$ SOHO/EIT \citep{delab1995} observe
the full-disk Sun with a cadence of 12~min and pixel resolution of
2.5$^{\prime\prime}$. It observes in four spectral bands centered on
Fe IX/X (171 \AA), Fe XII (195 \AA), Fe XV (284 \AA) and and He II
(304 \AA). For our current study, we used 171 \AA\ data.

\noindent $\bf {X-Ray~ Data:}$
To understand the evolution of flares and associated surges, we reconstructed
X-ray images from the Reuven Ramaty High-Energy Solar Spectroscopic
Imager (RHESSI; \citet{Lin02}). We reconstructed the images in
6-12 keV energy band from collimators (3F to 9F) using
the {\it CLEAN} algorithm, which has a spatial resolution
of $\approx 7''$ \citep{Hurford02}.

\section{MULTIWAVELENGTH OBSERVATIONS OF RECURRENT SURGES AND ASSOCIATED FLARES}
\label{multi}


The multiwavelength evolution of the observed solar surges and their
association with flares are described in following subsections.

\subsection{Temporal variations of  H$\alpha$ surges and flares }

\begin{figure}
\centering{
\includegraphics[width=6cm]{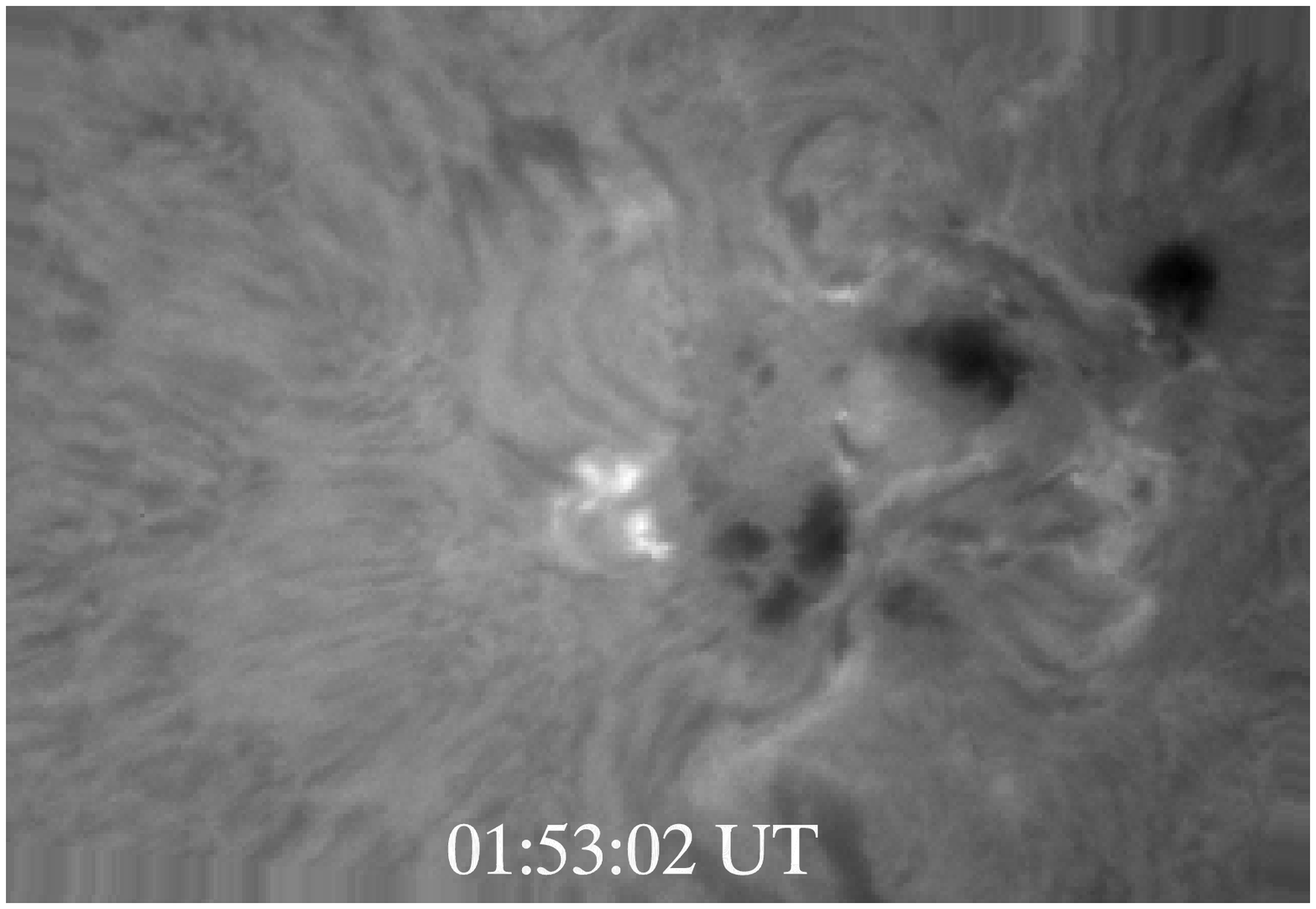}
$ \color{white} \put(-125,58){\vector(1,0){15}} \color{white} \put(-160,55){Surge 1}$
$ \color{white} \put(-98,70){\vector(0,-1){15}} \color{white} \put(-110,75){Flare 1}$
\hspace*{-0.45cm}
\includegraphics[width=6cm]{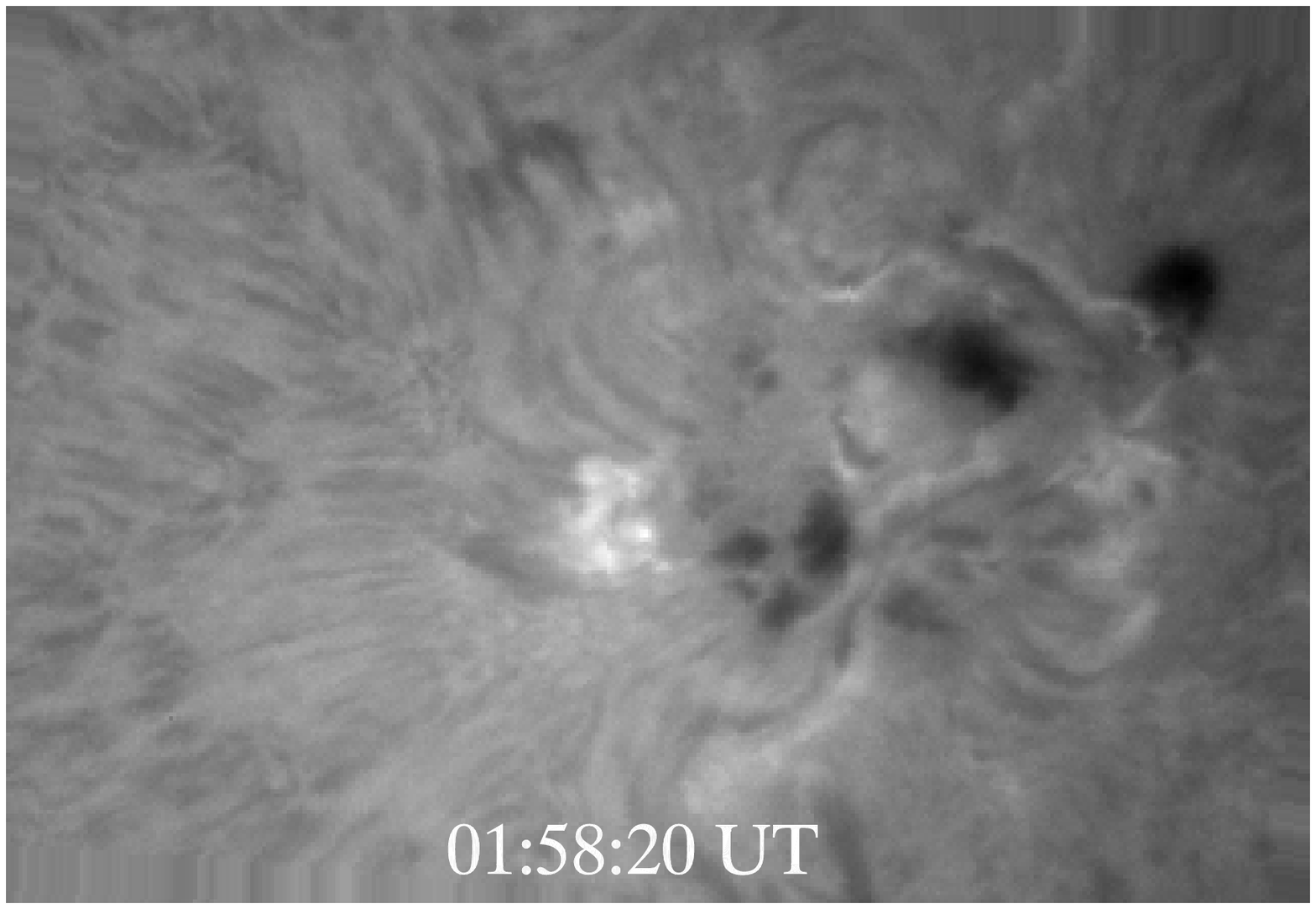}
\thicklines
}
\vspace*{-0.15cm}
\centering{
\hspace*{-0.09cm}
\includegraphics[width=6cm]{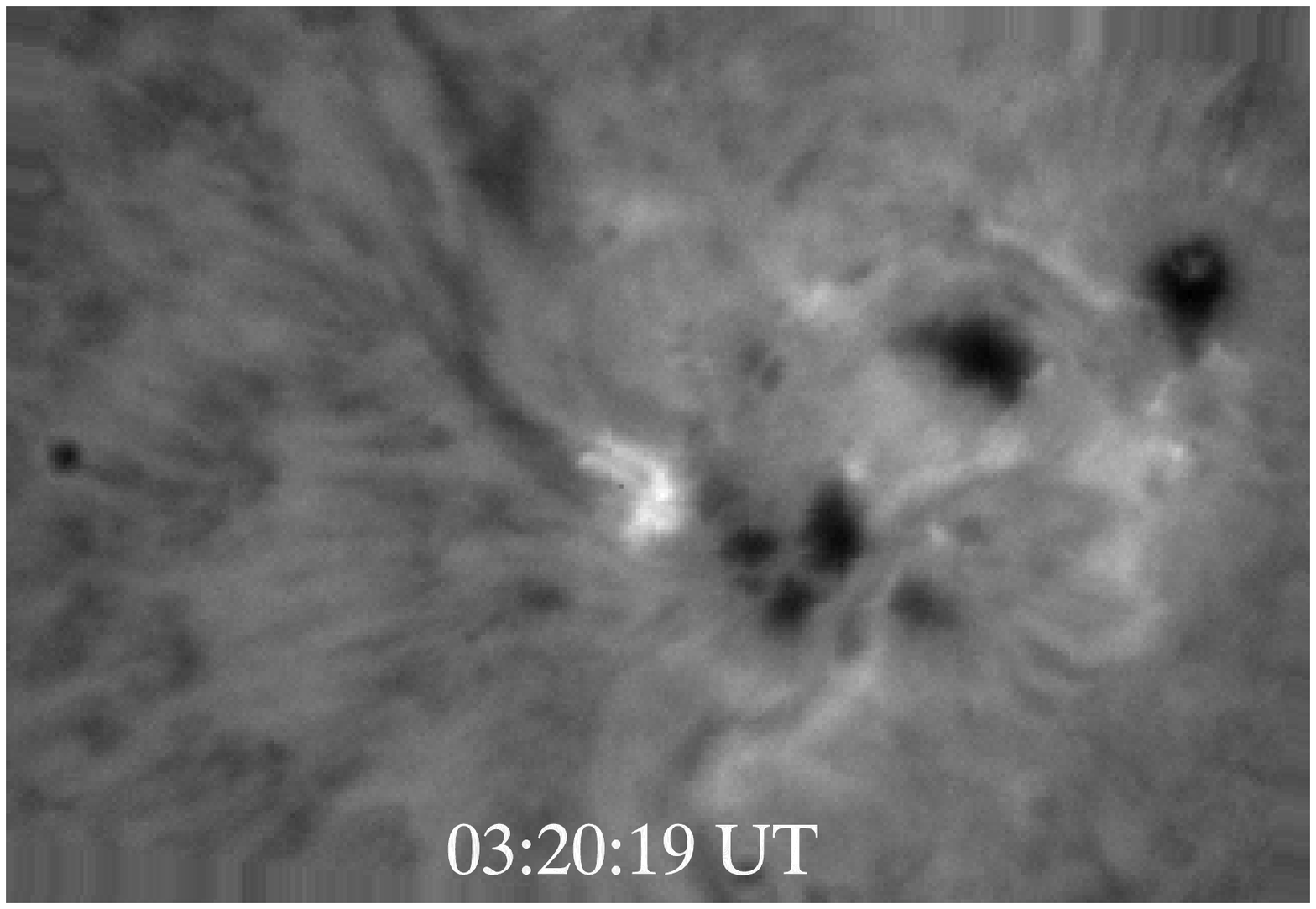}
$ \color{white} \put(-128,65){\vector(1,0){15}} \color{white} \put(-165,65){Surge 2}$
$ \color{white} \put(-95,70){\vector(0,-1){15}} \color{white} \put(-110,75){Flare 2}$
\hspace*{-0.45cm}
\includegraphics[width=6cm]{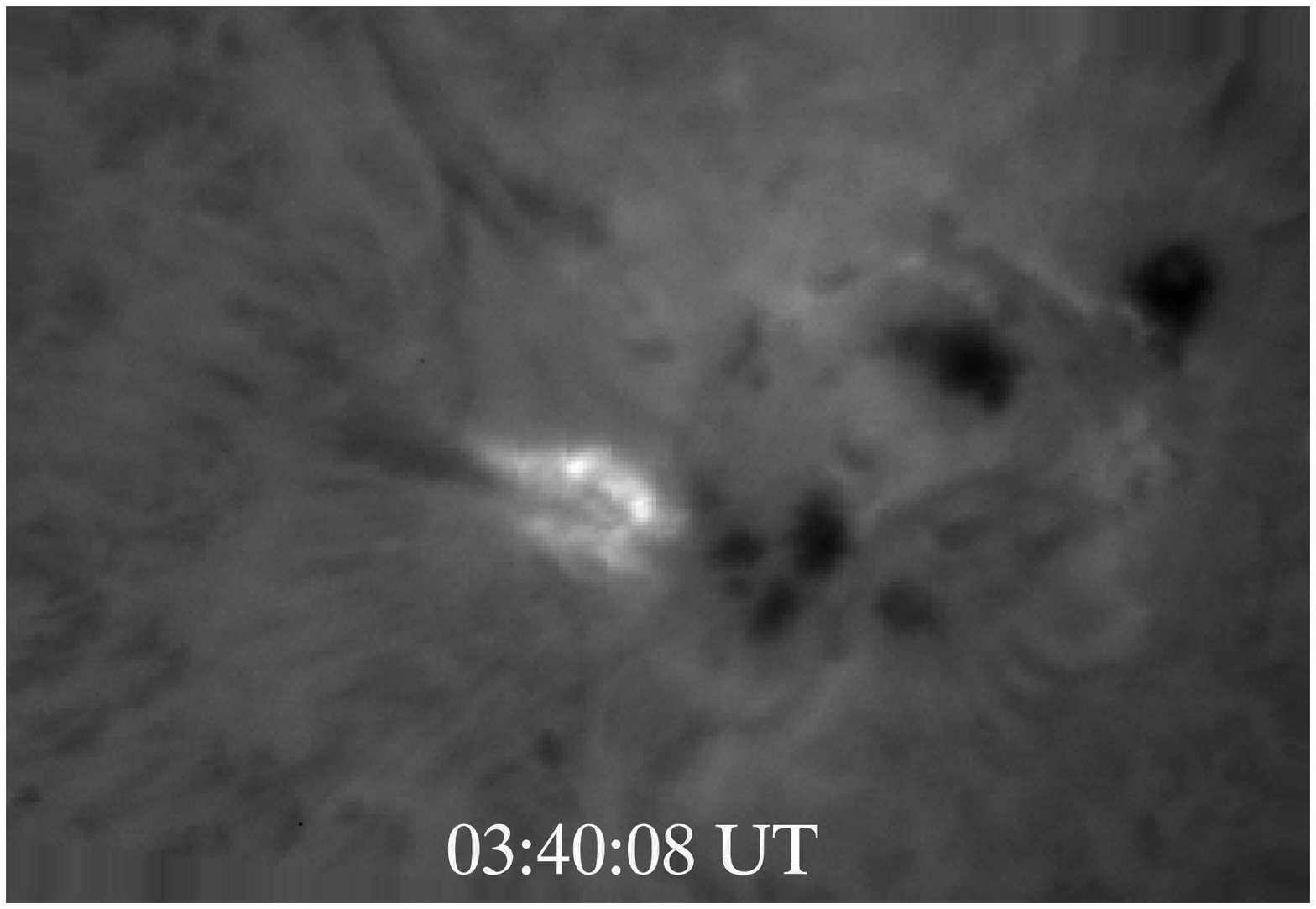}
$ \color{white} \put(-95,70){\vector(0,-1){15}} \color{white} \put(-110,75){Flare 3}$
$ \color{white} \put(-140,50){\vector(1,1){10}} \color{white} \put(-165,40){Surge 3}$
\thicklines
}
\centering{
\hspace*{-0.05cm}
\includegraphics[width=6cm]{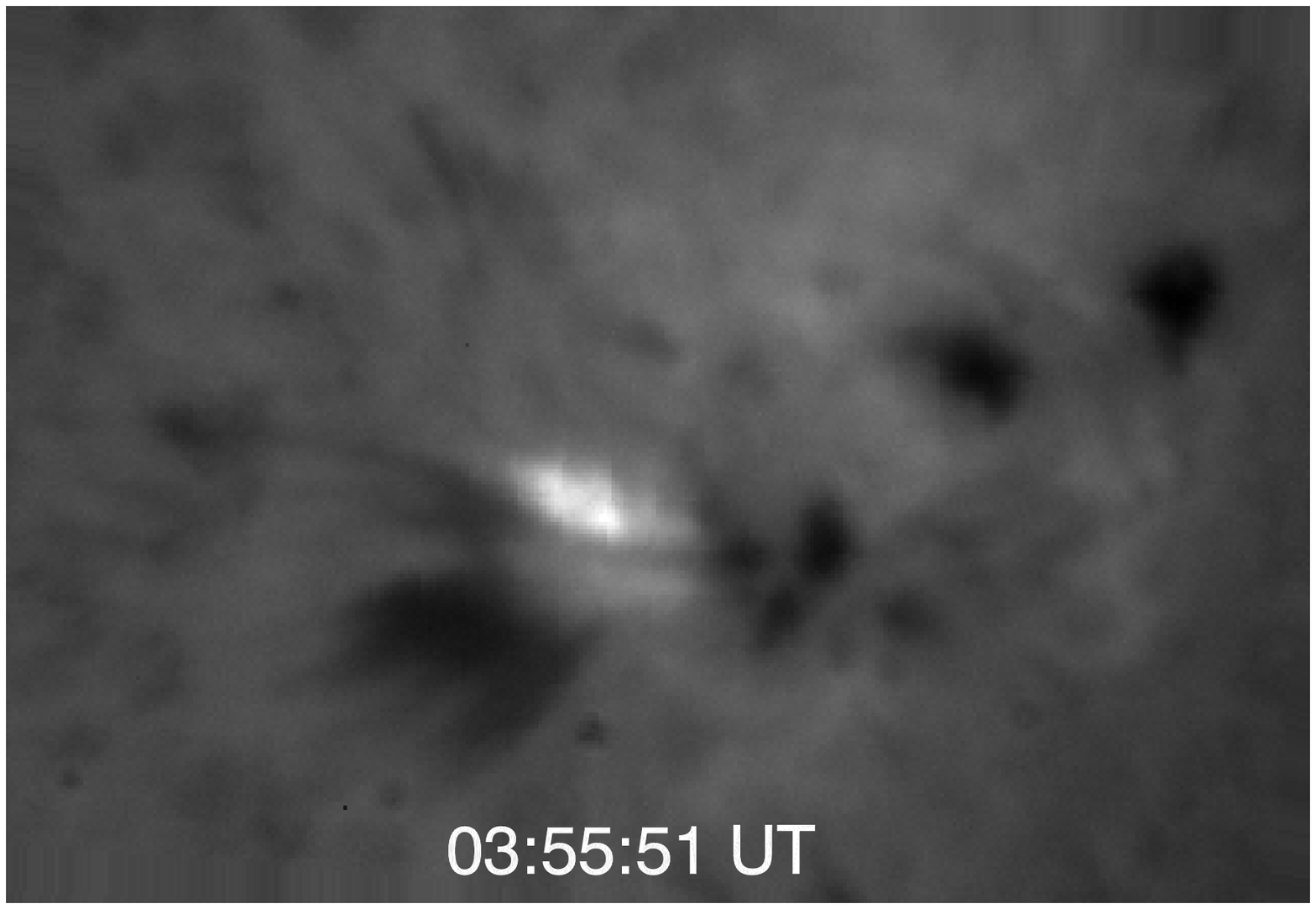}
$ \color{white} \put(-95,70){\vector(0,-1){15}} \color{white} \put(-110,75){Flare 3}$
$ \color{white} \put(-135,70){\vector(1,-1){10}} \color{white} \put(-165,70){Surge 4}$
$ \color{white} \put(-145,50){\vector(1,-1){10}} \color{white} \put(-175,52){Surge 5}$
$ \color{white} \put(-138,20){\vector(1,0){10}} \color{white} \put(-175,12){Surge 6}$
\hspace*{-0.75cm}
\includegraphics[width=6cm]{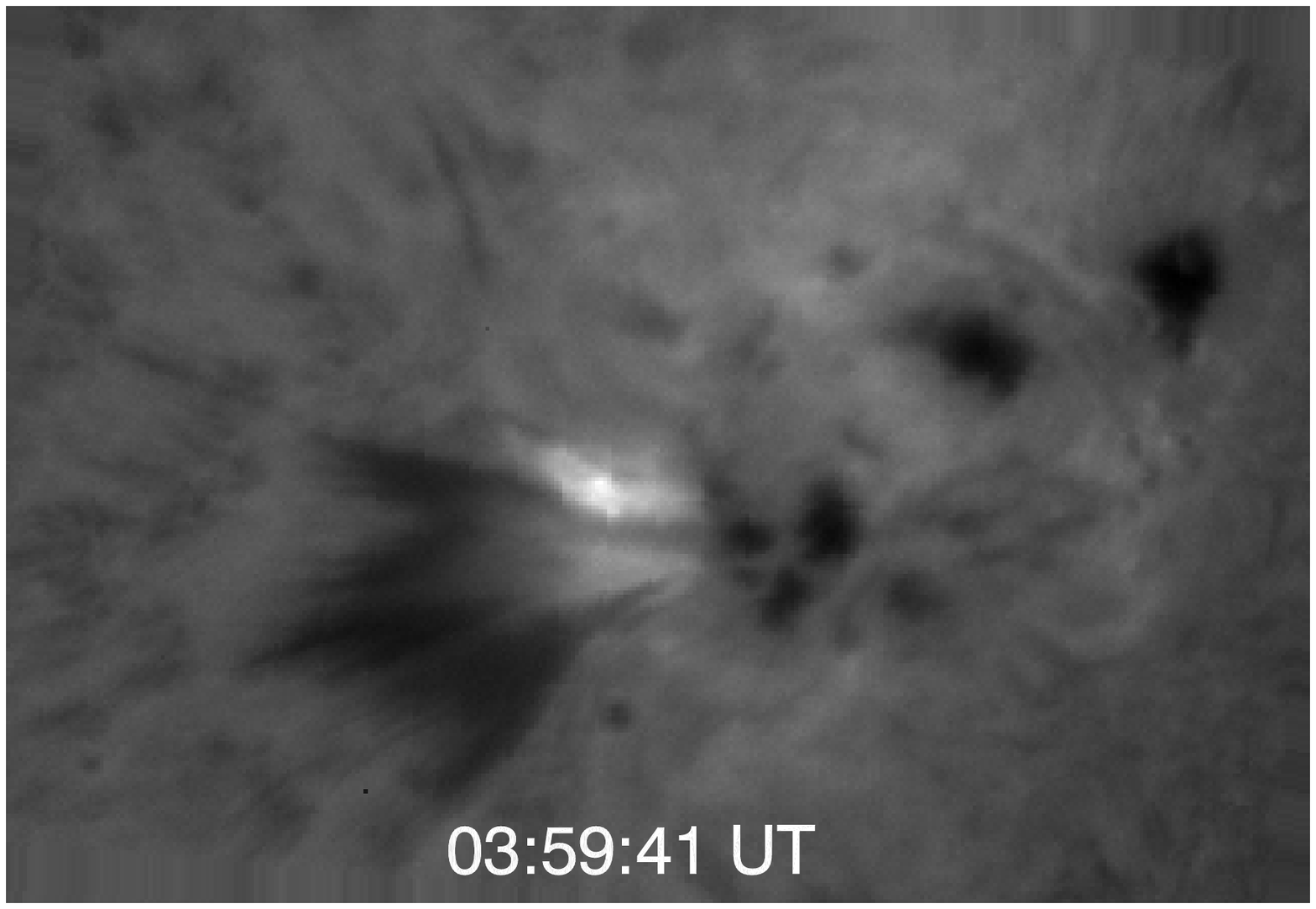}
$ \color{red} \put(-162,15){\vector(1,1){15}} \color{red} \put(-175,10){Leading edge}$
$ \color{red} \put(-81,65){\vector(0,-1){15}} \color{red} \put(-110,70){Reference foot-point}$
\thicklines
}
\centering{
\hspace*{-0.10cm}
\includegraphics[width=6cm]{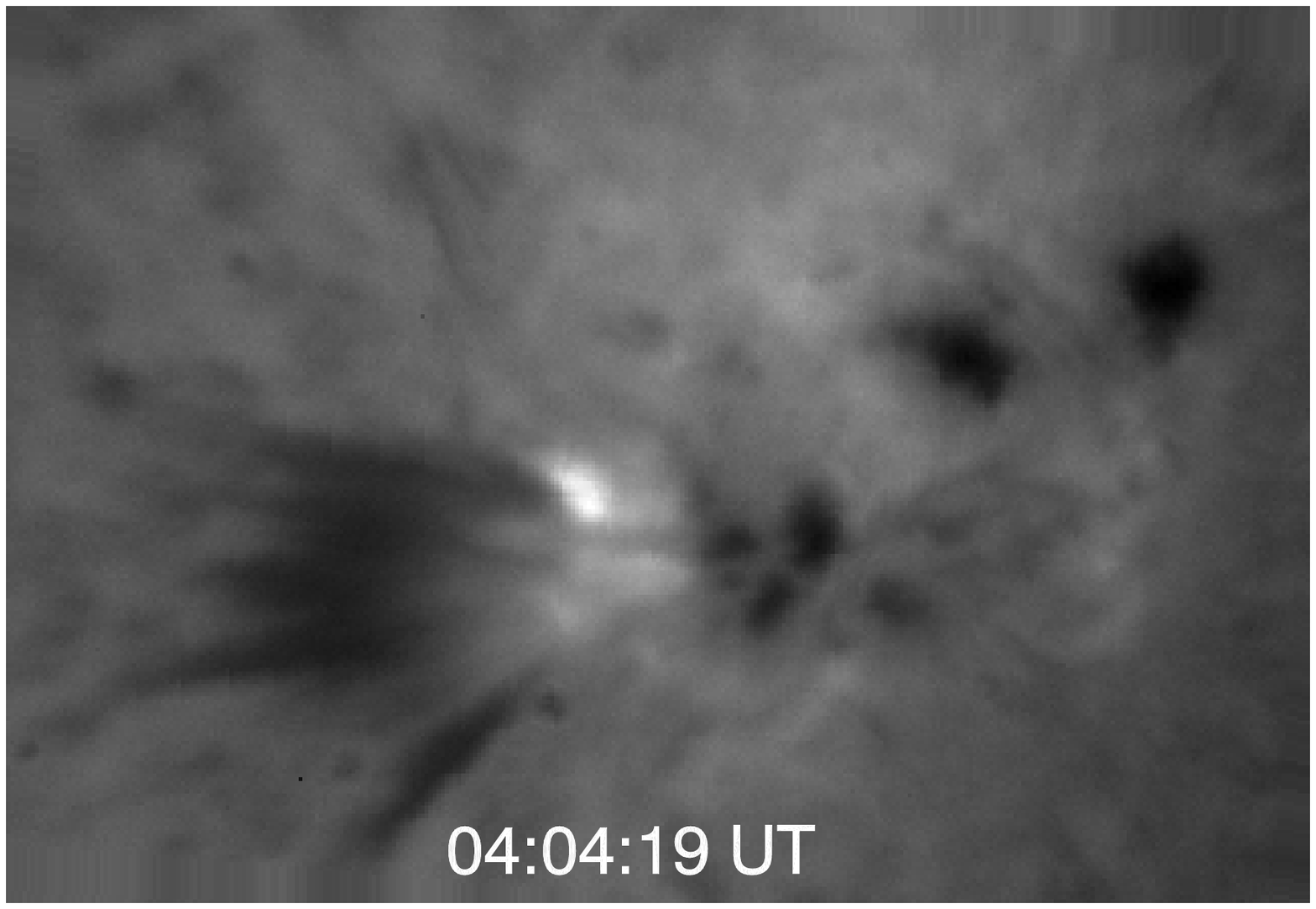}
\hspace*{-0.20cm}
\includegraphics[width=6cm]{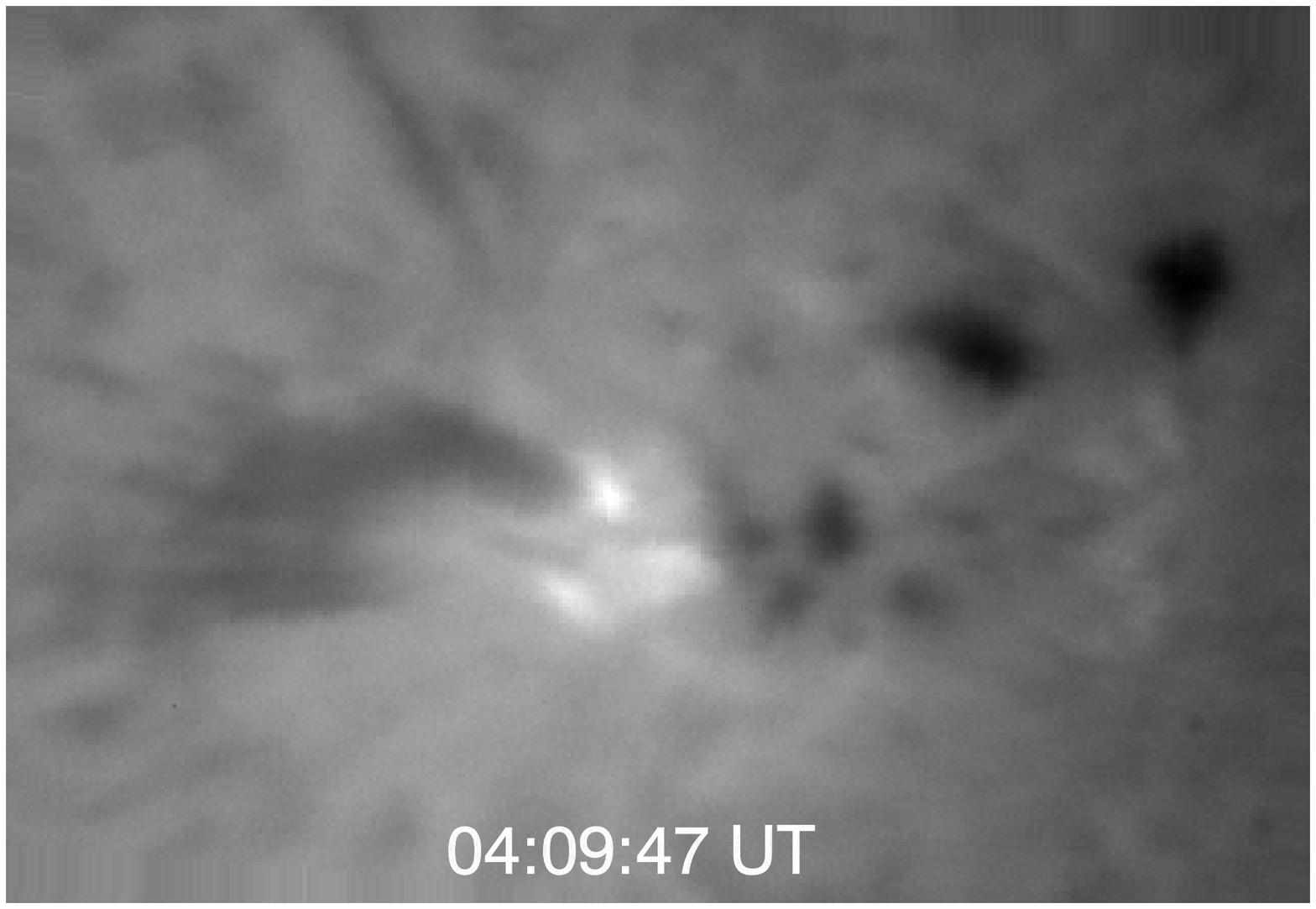}
$ \color{white} \put(-140,60){\vector(1,0){15}} \color{white}
\put(-175,60){Surge 7}$ } \caption{H$\alpha$ image sequence showing
the recurrent flare/surge activities on 25 October, 2003 in  AR 10484.
The field-of-view of each image is
320$^{\prime\prime}$$\times$200$^{\prime\prime}$.}
 \label{fig2}
\end{figure}

The ARIES H$\alpha$ images  during the surge activities on
25 October, 2003 from NOAA AR 10484, are presented  in Fig.~\ref{fig2}.
They 
show the dynamic evolution of the
recurrent surge activity from 01:50~UT to 04:15~UT. During the above
mentioned time period, we observed several  surges in H$\alpha$. 
The surge activity occurred in the following satellite sunspots of
the active region NOAA AR 10484.

\begin{figure}
\centering
\includegraphics[width=10cm]{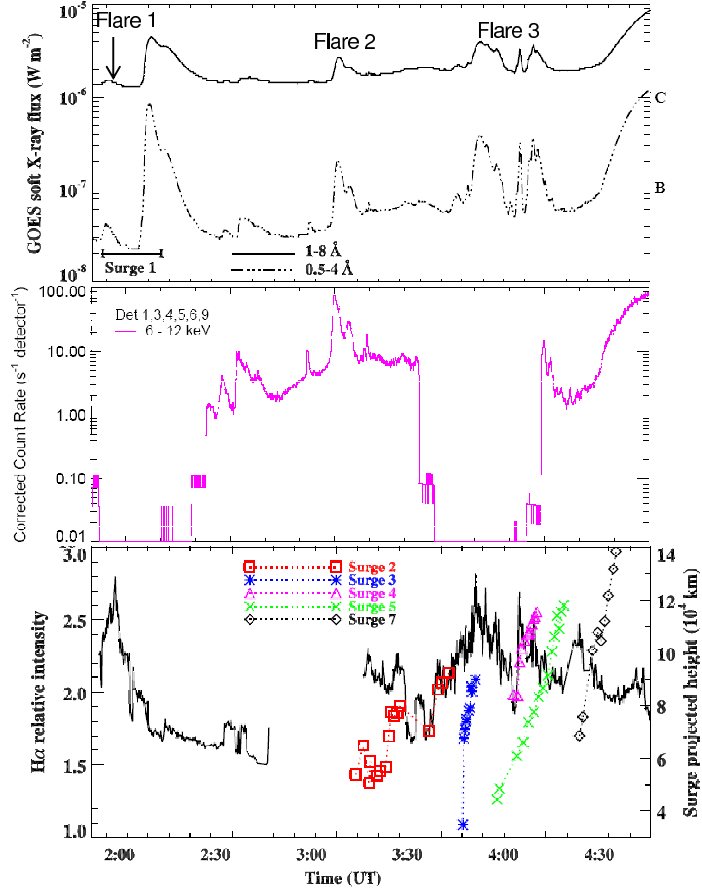}
\caption[] {\label{fig1} GOES Soft X-ray flux (top panel) profiles
in two different wavelengths on 25 October 2003. The H$\alpha$
relative intensity profile is well correlated with the GOES flux
(bottom panel). There was H$\alpha$ data gap in between 02:40--03:05
UT. There was recurrent small surge activities associated with
flares at $\sim$01:55 and between  03:00 UT and 04:30 UT (Flares 1, 2,
3).  The C4.3 flare during 02:02--02:12 was occurred in another AR
NOAA 10486 located at the eastern limb.
The length-time plot of the surges is shown and demonstrate the
surge-flare relationship. 
}
\end{figure}

Seven surges (Surge 1 to Surge 7)  were identified (Fig.~\ref{fig2}, see the arrows). Four of them were clearly associated
with H$\alpha$ brightenings.
To investigate in more detail the surge evolution and
the H$\alpha$ brightenings at their footpoints, we computed 
the H$\alpha$ relative intensity profile  of the brightenings, the time of  the H$\alpha$ 
 brightening  maxima, the onsets  of surges, and the time where the surge vanishes (Fig.~\ref{fig1} and Table 1).

In Fig.~\ref{fig1}, we also present the GOES soft X-ray light curve  obtained by full disk integration  (top panel) and the RHESSI satellite
thermal emission (6-12 keV) of the AR 10484 (middle panel). The  latter curve has unfortunately large time gaps,
however, we are able to identify 
two flares at $\sim$3:00 UT and at $\sim$4:00 UT. By comparing the three light curves of the Fig.~\ref{fig1},   
we conclude that all the C class X-ray flares of the GOES curve  (Flares 1, 2, and 3 with 3 bumps) 
occur in AR 10484, excepted one at 02:02 UT which is observed in the neighbouring  AR 10486.
We present in Table 1 the class of the flares.
We reconstructed images of RHESSI in the low energy band (6--12 keV) where there are enough counts.

To understand the location of the  X-ray sources with
H$\alpha$ brightenings, we co-aligned the H$\alpha$ data with the RHESSI thermal
source. 
The contours of the
X-ray sources is  drawn   in  two H$\alpha$ images  shown in Fig.~3. 
The location of the X-ray sources coincides with the
H$\alpha$ brightened footpoint for the two surges corresponding to the two flares observed by RHESSI (Flare 2, and Flare 3
with three bumps). 
It demonstrates the co-spatiality   of  the thermal sources and the H$\alpha$ brightenings.

\begin{table}[t]
\caption{Details of the surges and flares}
\small
\begin{center}
\begin{tabular}{|c|c|c|c|c|c|c|c|c|}
\hline
Name of & Surge & H$\alpha$ max  & Max & Flare & No. of & Max & Max Surge & Type III \\
surge & onset & Intensity & GOES & class & Flare & RHESSI & Length & \\
\hline
1 & 1:50 & 1:50 & 1:57 & C1.2 & 1 & No & 2:10 & 1:55-58\\
2 & 3:05 & $<$3:09 &  3:00 & C2.6 & 2 & 3:00 & 3:30 & 3:00 \\
3 & 3:36 & 3:30 -39 & 3:35-3:39 & C3.9 & 3 (bump1) & No & 3:40 & 3:35-3:45 \\
4 & 3:50 & 3:52 & 3:52 & C3.4 & 3(bump2) & No & 4:00 & --  \\
5 & 3:42 & 3:40 & 3:40 & C3.9 & 3(bump1) & No & 4:05 & 3:35-3:45\\
6 & $<$3:55 & -- &-- &-- &-- & No & 4:09 & --  \\
7 & 4:09 & 4:10 & 3:57 & C3.6 & 3(bump3)& $<$4:00 & 4:20 & 4:20 \\
\hline
\end{tabular}
\end{center}
\label{default}
\end{table}

\subsection{Spatial variations of  H$\alpha$ surges}
We describe in details each surge  shown in Fig.~\ref{fig2}  and give quantitative results in Table 1.
We have calculated  for each surge its  extension versus time 
 from the  brightened footpoint to  its
leading edge and its speed.    These results are shown in Fig.~\ref{fig1} (bottom panel) and in Table 1.

 Surge 1 was a small ejection of plasma in
the North-East direction, and was associated with 
Flare 1 and an H$\alpha$ brightening (Fig.~\ref{fig2}).  Due to unavailability of
high-resolution H$\alpha$ observations, we could not see its
evolution in details.
 Surge 2 started around 03:06 UT in North-East direction, increased within two periods and finally
decayed around 03:30 UT. This surge
was associated with Flare 2 ( C2.9 ) as well as  with  an H$\alpha$ brightening at its footpoint. 
Surge 3 started very impulsively around 03:36 UT with a rapid
enhancement of an H$\alpha$ brightening (cf., Fig.~\ref{fig1}).
The plasma was ejected in North-East direction, however, slightly
East in comparison to the direction of the previous surges. This
surge decayed in 4 minutes around 03:40 UT. The speed of this surge expansion is estimated to 200 km/s,
which is the faster.
When the H$\alpha$ brightening reached its maximum, the surge was
decaying.  It could be associated with the first bump of Flare 3 (Flare 3, bump 1). 
Similarly, Surge 4 started to erupt around 03:45 UT in the
North-East direction from the triggering site of the recurrent
surges  during the impulsive and maximum phase of the H$\alpha$
brightening.

At 03:55 UT, co-temporally to Surge 4, Surges 5 and 6  have already
started to erupt in the East and South-East direction. Surge 5 is
bigger than Surge 6, and more elongated in size. After some time,
they appeared as a single surge forming together a fan-type dark
structure.   Surge 5  and 6 were not associated with intense
brightenings at their footpoints. However, during the surge onset
period, we found faint brightenings.  Possibly the flare brightening
may be hidden by the dark surge material or it is associated with the first bump of Flare 3.1.
 Thereafter, they merge with each other,
and individual surges were no longer resolved. We could only
calculate the length of Surge 5 as the contrast of surge 6 was comparatively low.
 The length-time plot of Surge 5  shows that the surge erupt
after Flare 3. Surge 7 started around 04:09
UT again in the East direction, and was initiated from a H$\alpha$
brightening corresponding to Flare 3.3.  Surge 7 reached the largest  height compared with the
other surges. The maximum projected length in case of Surge 7 is
around 140 Mm and its speed close to 100 km/s.

We conclude that the surge
onsets were commonly associated with the impulsive phase of the GOES
flares  and reached their maximum length during  the
decay of H$\alpha$ brightenings. Their velocities are between 25 to 200 km/s.

\subsection{SOHO/EIT observations and topology analysis}
To
compare the H$\alpha$ brightenings and the coronal brightenings visible in
EIT, as well as surges, we overlaid EIT contours over  the H$\alpha$
images at a four instances. The results are shown in Fig.~4. 
At 03:24 UT,  at 03:48 and 04:12 UT, the EIT bright regions are elongated structures and closely
co-spatial with the H$\alpha$ surge footpoint brightenings even they did not correspond to the onsets of the flares.
The heating indicated by the brightenings in multi-wavelengths is local and nearly co-spatial.  The bright H$\alpha$ brightenings are possibly  parts of the
surges, as suggested previously \cite{schmieder84} by looking at the
Doppler shifts of such  structures, which show  a continuity in the Doppler-shift pattern between bright and dark structures.
We would need to have informations on the Doppler
shifts all along the structures to confirm that bright and dark
bundles belong to  a unique structure.

An interesting observation is the shape of the loops visible in EIT before the surge activity.
Fig.~5 displays the SOHO/EIT 171 \AA\ image at 01:00:14
UT.
The dotted $'$V$'$ shaped line
in the East direction indicates a fan-shape structure of magnetic field lines, which is
similar to the fan-shape of the bunches of surges 4, 5, and 6.
All the H$\alpha$ surges erupted in the same cone  direction. 


A question arises: Are the field lines inside the fan open or closed?
We used the Potential Field Source Surface (PFSS) extrapolation
code to derive the global structure of the Sun on October 25, 2003.
Fig.~\ref{fig4-bis} shows that the bunches of field lines in the fan are in fact closed field lines linking the two active regions: AR 10484 and AR 10486. 

In Fig.~4, the  H$\alpha$ image at 03:59 UT is overlaid
by MDI contours. From this snapshot, it is shown that  the
footpoints of the surges are in a  mixed polarity region,  dominated
by the positive polarity but the edge of a  negative main  polarity spot of the region. The region of the ejection of the surges is 
located between magnetic field lines on the East side of the active
region and the active region itself overlaid by close loops. It
means that there may be separatrix or Quasi-separatrix layers (QSLs)
in that region. Field lines change easily of connectivity according to photospheric motions of the mixed polarities. It is a favourable region for reconnection. We
suggest that the reconnection is done successively in different
locations along the separatrix from a northern  point to southern
one. The material of the surge would escape along the pre-existing
field lines, which make a fan-shape structure similar to the EIT loops.
The reconnection would heat locally the plasma in these reconnected field lines up to 1 MK.
These events are transients and may participate to the global coronal heating only  with a minor effect.

\subsection{Radio Observations}

Fig.~\ref{fig4-ter} (bottom-panel) shows the dynamic radio spectrum of
Learmonth observatory in the frequency range 25-180 MHz. During the
maximum phase of recurrent surge activities, a series of Type III
bursts has been observed (Table 1). We noticed two Type III bursts during the
maximum of  Surge 1 before 02:00 UT. During Surge 2, two sets of
Type III bursts are evident, respectively around 03:30 UT and 03:40
UT. Type III radio bursts generally represent the escaping electrons
along open field lines \citep{dulk79,nitta08}. The appearance of
recurrent Type III bursts during the peak evolutionary phases of the
surges, may be most likely the signature of the generation of
non-thermal particles due to reconnection episodes. It may be most
plausible during the initiation of recurrent surge activity and the
eruption of surge material that the magnetic field lines come closer
and reconnect. Due to the reconnection, the energetic particles are
produced and hence we observe the series of Type III radio bursts.
Therefore, the presence of recurrent Type III bursts provide an
evidence of multiple magnetic reconnections at the activity site,
and bulk plasma acceleration along the  field lines forming the
surges.  
No Type III bursts have been identified in WIND satellite data as have shown other authors for a similar event \citep{bent2000}.
Therefore it confirms that the accelerated particles are reaching only to the low corona, 
as suggested by the radio frequency range 
($\sim$0.2, 0.3 solar radii). These particles are certainly following the system of large loops existing between the two active regions already mentioned in the previous 
section. 
These transient  cool jets and associated particles are confined in  close loops and 
cannot play an  important role in the outflows observed  commonly at the periphery of active regions in open field and definitively not participate to
the solar wind.



\section{MAGNETIC FIELD ACTIVITY RELATED WITH RECURRENT SURGES}
\label{mag}

In order to understand the relation between the recurrent surge
activity and the  magnetic field configuration of the region,  we
present the evolution of magnetic field before, during and after the
surges (Fig.~\ref{mdi_evo}). From MDI time-series data, we
see that a large  bipole emerged ahead of a remnant active region as
it crosses the limb. On the 18-20 October, the two main polarities
of the emerging bipoles have a ``tongue-shape'' as it is
frequently observed during strong emergence \citep{Chandra09}. The
polarities of the old active region broke in many segments in the
surrounding of the bipole and created the $\delta$ configuration of
the AR 10484. This is evident as largely extended negative polarity
area surrounded by positive polarities in the active region. Already
on October 22, the polarities started  to disperse and the active
region entered in a decay phase. Fig.~\ref{mdi_evo} (top) shows
the MDI magnetogram on 24 October 2003 at 14:23 UT before the surge
activity. Opposite polarities surrounding the main polarities escape
radially away from the spots. Around the left/right
(negative/positive) main polarity, there is  a ring of opposite
polarities (positive/negative) respectively. Such a ring is observed
successively on October 22 and on 24 October at 14:23 UT. This is
the scenario of the building up of magnetic activity and complexity
in and around the origin site of recurrent solar surges in AR 10484.

During the decay phase of  sunspots, there is a radial dispersion of
the flux with radial ``sea serpent'' structures with positive and
negative polarities forming a ring of polarities that progressively
reaches the network. This region surrounding the spot is called moat
region \citep{Harvey73}. The polarities that are present in this
moat region are  in the dynamics of the dispersion. Looking at the
MDI time-series data, we see that this ring of polarities appear several times
during the dispersion phase of the magnetic fluxes. During this
time, some small flux were escaping from the negative/positive main
polarity and {\bf most likely} cancel with the moat polarities. This phenomena is a
common process to decrease the flux during the decaying phase of an
active region. The enlarged part of the active region, where the
surges were originated, has been displayed in Fig.~\ref{mdi_evo}
(bottom).

We notice that the positive polarity P0 is submitted to the ambient
dispersion process. P0 broke into three pieces i.e. P2, P3, and P4 on 25
October at 01:35 UT. From the main negative polarity, a small
negative polarity N1 escaped and disappeared at 06:23 UT on 25
October. This negative polarity (N1) was being  cancelled with parts
of small P2 and P3 polarities. The cancellation of P0 was done in
successive annihilations of positive and negative fluxes during the
spatial expansion of main polarities. This annihilation occurs very
close to the inversion line in close proximity of P2, P3, P4, N1,
and the main negative polarity. We also followed the evolution of P1
and N0 polarities with time and found that these polarities are
increasing in size as well in magnitude, and coming closer to each
other for cancellation.

We computed the variation of magnetic fluxes in the region of these
surges. For this purpose, we selected a box as shown in Fig.~\ref{mdi_evo}. 
The temporal variation of positive, negative and
total fluxes before, during and after the surge activity inside the
box is shown in Fig.~\ref{flux}. The plot shows that the positive
flux is increasing before the surge activity due to the  emergence
of positive ring polarities. On the other hand during the surge
activity (shown by vertical line in Fig.~\ref{flux}), we found the
decrement in positive, negative, and total flux. The constant
decreasing negative flux demonstrates that the active region is in a
decay phase. The decrease of total  flux at the bottom of surges
indicates that the flux cancellation is on going process, which would
be important factor for the triggering of solar surges. However it
is difficult to detect  the decrease of the negative flux associated
with  the cancellation field at the base of the surge because the
decrease of such a flux is one order of magnitude less according to
the decrement of positive flux. 

Under the base-line of our interpretation 
based on morphological investigation of MDI data as well as
temporal variation of the magnetic fluxes, it 
may also be quite possible that the flux cancellation could be a signature of
submergence of the magnetic polarities.
Therefore, the energy released
by reconnection in the low solar atmosphere would be transformed into
thermal heating and kinetic energy.
These underlying processes
of surges are  starting again when  the   ring of opposite polarities
emerges around the main spots. An other important reason that such
cancellation leads to  surge activity, is the existence of  the QSLs , region where field lines can change of connectivity.

In the present study
under the baseline of our observations, we found the emergence and
fragmentation of  magnetic polarities, and thereafter their
cancellation with the main magnetic polarities that trigger the
recurrent surges and associated energy release  (heating) at their footpoints.
Fig.~\ref{cartoon} displays the schematic cartoon showing the
surge association due to the reconnection of the  neighbouring  field lines
(i.e. positive) with the opposite polarity field region. The ``X''
symbol shows the reconnection point in between opposite polarity
field regions.

\section{DISCUSSION AND CONCLUSIONS}
\label{discussion}
In the present paper, we outline a multiwavelength study of seven
recurrent surges and associated flare brightenings on  October 25
2003 from the active region NOAA AR 10484. For this study, we use
ARIES H$\alpha$, SOHO/EIT, MDI, and RHESSI observations. Out of
seven surges, five surges were associated with  footpoint
brightenings and corresponding flare energy release. In case of
surge five and six, we could not notice footpoint brightening. One
possibility to not see footpoint brightening is that these surges
are very dynamic and extended, as well as due to the large area of
these surges the compact footpoint brightening may be hidden in the
cool plasma eruption. The length-time plots of brightenings
associated surges indicate that the length is increasing or
sometimes decreasing (as in the case of surge 2) with the increment
or decrement of the brightness at footpoint. This is the evidence
that compact energy release at the base is pumping the plasma
material in the form of surge. All the surges ejected in the North-East
to South-East direction.
The speed of the surges is between 25  and 200 km/s. This kind of large velocity has been
observed in spicules Type II  in Ca H as reported by \cite{barte10}.  They claimed that they can follow the up flows in 
hot structures (1 to  2 MK) which could  participate to the coronal heating.
We cannot confirm their findings even if hot material (1MK) is observed.
Recently, the hot plasma up flows
were also detected near the boundary of active regions, which may be
potential candidates for the coronal heating \citep{Bart11}. Such
type of jets, surges, and confined plasma dynamics
have been modelled recently in form of the evolution of slow shocks
that could carry the hot plasma material followed by the cool
plasma in the under pressure region in  magnetic flux tubes
\citep{Sri11,Sri12,Pra12}. Afterall, in the present
observations, the hot plasma is detected at the base of the cool
jets and not at the top. The shock mechanism cannot be at work in
the present situation.
Both hot and cool plasma follow field lines belonging to the same system joining two active regions.
This plasma is not in open field lines and therefore, will not implement the commonly 
observed  outflows in the outskirts of the active regions
which  could participate to the initiation of the solar wind  \citep{harra08,delzanna08}.

The MDI magnetograms reveal positive flux of opposite polarities
around the main spots. This behaviour could indicate that the active
region is in a decay phase with the presence of a moat region. The
polarities are very much fragmented in this moat region. The
dispersion motion pushed the  parasitic positive polarities to merge
with negative polarities in the moat region. The positive parasitic
polarity changes shape, breaking into pieces and cancel with
neighbouring  opposite polarities during the dispersion of the
dipoles of the moat. The cancellation of flux initiates the surges.
The horizontal motion of the polarities in the moat region is
important to push the opposite polarities together and forces the
reconnection. The field lines of the sunspot penumbra are
reconnecting with the pre-existing open field external to the active
region. It is frequently observed that surges occur at the periphery
of active region close to parasitic polarities \citep{kuro2007, Brooks07}. MHD
simulations show that surges and jets occur when polarities merge
close to open field lines like in coronal holes \citep{Moreno08}. A
question subsists if the surges are produced by squeezing and
compression of the plasma or by reconnection \citep{Isobe07}. Flux
cancellation at the bottom of the surges might give the power for
recurrent surge activities. This scenario is supported by
\citet{yokoyama1996} on the basis of numerical simulations. The
flares that formed nearby the footpoints of the surges and the
Type-III radio bursts that were observed during these events are
evidences of magnetic reconnections \citep{shibata1994,canfield96,bent2000}.
The presence of open field close to the active region  favours the
occurrence of surges and jets \citep{Moreno08}.

However, the formation of various types of jets (e.g., surges,
coronal jets, spicules) at different spatio-temporal scales can
either be driven by direct reconnection process , or sometimes
reconnection triggered them indirectly. This depends upon the height
of the reconnection site as well as the local plasma and magnetic
field conditions. \cite{Ster93} have demonstrated that the pressure
pulse generated by energy release in the non-magnetized atmosphere
can trigger the surge material. Recently, the cool jets and long
spicules have been reported as driven by the reconnection generated
velocity pulses sufficiently evolved in the chromosphere
\citep{Sri11, Mur11, Pra12}. However, such magnetohydrodynamic
pulses and their steepening in form of localized shocks were
associated mostly with non-flaring regions, where they attain
sufficient spatio-temporal scales and ambient plasma conditions for
their growth and subsequently the triggering of the various types of
jets. Moreover, such types of pulse driven jets exhibit
quasi-periodic rise and fall of the plasma material. In the present
observations, we get the recurrent surge jets and associated flare
energy release. Photospheric magnetic field cancellation due to the
fragmentation of the emerging opposite polarities and associated
compact flare energy release are clearly evident, which may further
release the plasma along the open field lines in the form of recurrent
surges. Therefore, the recurrent flare energy release at the base of
the surge locations near their footpoints generated due to
photospheric reconnection may be the main cause for the surge
eruptions. Collision of the magnetic polarities triggers the compact
flares and associated surges. Therefore, we rule out the other causes
e.g., the generation of the pressure or velocity pulses in the surge
region, evidence of some explosive event \citep{Maj09}, role of the
twist \citep{pariat10} etc for the presented observations of the jet
formation.

In conclusion, we report on a multiwavelength observational study of
energy build-up and dynamics in the form of multiple surge eruptions
associated with flares due to successive reconnections initiated by
magnetic flux cancellations. However, the future multiwavelength
observations and related MHD modeling should be carried out using
recent high spatial and temporal resolution observations from space
(e.g., SDO, Hinode) and complementary ground-based observations to
understand the  initiation, energetics, magnetic field topology and
dynamics of surges.


\acknowledgments
We acknowledge the valuable suggestions of the referee
that improved our manuscript considerably.
The authors thank CEFIPRA Project 3704-1 for its support to this
study on ``Transient events in Sun Earth System'' during our
bilateral collaboration. We acknowledge the space borne instruments onboard
SOHO, RHESSI, and ground based Learmonth, Australia for the data
used in this study. SOHO is an international cooperation between ESA
and NASA. BS thanks the ISSI group  (Bern) lead by Klaus
Galgaard on Flux emergence which helps her to have a clear idea of
the development of this surge activity.
AKS acknowledges Shobhna Srivastava for her patient encouragement.
\bibliographystyle{apj}
\bibliography{references}

\begin{thebibliography}{}

\bibitem[\protect\citeauthoryear{{Beck} et~al.}{{Beck} et~al.}{2007}]{Beck07}
{Beck}, C., {Bellot Rubio}, L.~R., {Schlichenmaier}, R.,  \& {S{\"u}tterlin},
  P. 2007, \aap, 472, 607

\bibitem[\protect\citeauthoryear{{Bentley} et~al.}{{Bentley}
  et~al.}{2000}]{bent2000}
{Bentley}, R.~D., {Klein}, K.-L., {van Driel-Gesztelyi}, L., {D{\'e}moulin},
  P., {Trottet}, G., {Tassetto}, P.,  \& {Marty}, G. 2000, \solphys, 193, 227

\bibitem[\protect\citeauthoryear{{Brooks}, {Kurokawa}, \& {Berger}}{{Brooks}
  et~al.}{2007}]{Brooks07}
{Brooks}, D.~H., {Kurokawa}, H.,  \& {Berger}, T.~E. 2007, \apj, 656, 1197

\bibitem[\protect\citeauthoryear{{Brooks}, {Ugarte-Urra}, \& {Warren}}{{Brooks}
  et~al.}{2008}]{brooks08}
{Brooks}, D.~H., {Ugarte-Urra}, I.,  \& {Warren}, H.~P. 2008, \apjl, 689, L77

\bibitem[\protect\citeauthoryear{{Canfield} et~al.}{{Canfield}
  et~al.}{1996a}]{Can96}
{Canfield}, R.~C., {Reardon}, K.~P., {Leka}, K.~D., {Shibata}, K., {Yokoyama},
  T.,  \& {Shimojo}, M. 1996a, \apj, 464, 1016

\bibitem[\protect\citeauthoryear{{Canfield} et~al.}{{Canfield}
  et~al.}{1996b}]{canfield96}
{Canfield}, R.~C., {Reardon}, K.~P., {Leka}, K.~D., {Shibata}, K., {Yokoyama},
  T.,  \& {Shimojo}, M. 1996b, \apj, 464, 1016

\bibitem[\protect\citeauthoryear{{Chae} et~al.}{{Chae} et~al.}{1999}]{Chae99}
{Chae}, J., {Qiu}, J., {Wang}, H.,  \& {Goode}, P.~R. 1999, \apjl, 513, L75

\bibitem[\protect\citeauthoryear{{Chandra} et~al.}{{Chandra}
  et~al.}{2006}]{chandra06}
{Chandra}, R., {Jain}, R., {Uddin}, W., {Yoshimura}, K., {Kosugi}, T., {Sakao},
  T., {Joshi}, A.,  \& {Deshpande}, M.~R. 2006, \solphys, 239, 239

\bibitem[\protect\citeauthoryear{{Chandra} et~al.}{{Chandra}
  et~al.}{2009}]{Chandra09}
{Chandra}, R., {Schmieder}, B., {Aulanier}, G.,  \& {Malherbe}, J.~M. 2009,
  \solphys, 258, 53

\bibitem[\protect\citeauthoryear{{De Pontieu} \& {McIntosh}}{{De Pontieu} \&
  {McIntosh}}{2010}]{barte10}
{De Pontieu}, B.,  \& {McIntosh}, S.~W. 2010, \apj, 722, 1013

\bibitem[\protect\citeauthoryear{{De Pontieu} et~al.}{{De Pontieu}
  et~al.}{2011}]{Bart11}
{De Pontieu}, B., et~al. 2011, Science, 331, 55

\bibitem[\protect\citeauthoryear{{Del Zanna}}{{Del Zanna}}{2008}]{delzanna08}
{Del Zanna}, G. 2008, \aap, 481, L49

\bibitem[\protect\citeauthoryear{{Delaboudini{\`e}re}
  et~al.}{{Delaboudini{\`e}re} et~al.}{1995}]{delab1995}
{Delaboudini{\`e}re}, J., et~al. 1995, \solphys, 162, 291

\bibitem[\protect\citeauthoryear{{Dulk}, {Melrose}, \& {Suzuki}}{{Dulk}
  et~al.}{1979}]{dulk79}
{Dulk}, G.~A., {Melrose}, D.~B.,  \& {Suzuki}, S. 1979, Proceedings of the
  Astronomical Society of Australia, 3, 375

\bibitem[\protect\citeauthoryear{{Engell} et~al.}{{Engell}
  et~al.}{2011}]{engell11}
{Engell}, A.~J., {Siarkowski}, M., {Gryciuk}, M., {Sylwester}, J., {Sylwester},
  B., {Golub}, L., {Korreck}, K.,  \& {Cirtain}, J. 2011, \apj, 726, 12

\bibitem[\protect\citeauthoryear{{Georgakilas}, {Koutchmy}, \&
  {Alissandrakis}}{{Georgakilas} et~al.}{1999}]{Gu99}
{Georgakilas}, A.~A., {Koutchmy}, S.,  \& {Alissandrakis}, C.~E. 1999, \aap,
  341, 610

\bibitem[\protect\citeauthoryear{{Harra} et~al.}{{Harra}
  et~al.}{2008}]{harra08}
{Harra}, L.~K., {Sakao}, T., {Mandrini}, C.~H., {Hara}, H., {Imada}, S.,
  {Young}, P.~R., {van Driel-Gesztelyi}, L.,  \& {Baker}, D. 2008, \apjl, 676,
  L147

\bibitem[\protect\citeauthoryear{{Harvey} \& {Harvey}}{{Harvey} \&
  {Harvey}}{1973}]{Harvey73}
{Harvey}, K.,  \& {Harvey}, J. 1973, \solphys, 28, 61

\bibitem[\protect\citeauthoryear{{Hurford} et~al.}{{Hurford}
  et~al.}{2002}]{Hurford02}
{Hurford}, G.~J., et~al. 2002, \solphys, 210, 61

\bibitem[\protect\citeauthoryear{{Isobe}, {Tripathi}, \& {Archontis}}{{Isobe}
  et~al.}{2007}]{Isobe07}
{Isobe}, H., {Tripathi}, D.,  \& {Archontis}, V. 2007, \apj, 657, L53

\bibitem[\protect\citeauthoryear{{Kayshap}, {Srivastava}, \&
  {Murawski}}{{Kayshap} et~al.}{2012}]{Pra12}
{Kayshap}, P., {Srivastava}, A.~K.,  \& {Murawski}, K. 2012, ApJ, submitted

\bibitem[\protect\citeauthoryear{{Kitiashvili} et~al.}{{Kitiashvili}
  et~al.}{2010}]{Kitiashvili10}
{Kitiashvili}, I.~N., {Bellot Rubio}, L.~R., {Kosovichev}, A.~G., {Mansour},
  N.~N., {Sainz Dalda}, A.,  \& {Wray}, A.~A. 2010, \apj, 716, L181

\bibitem[\protect\citeauthoryear{{Kubo} et~al.}{{Kubo} et~al.}{2008}]{Kubo08}
{Kubo}, M., {Lites}, B.~W., {Shimizu}, T.,  \& {Ichimoto}, K. 2008, \apj, 686,
  1447

\bibitem[\protect\citeauthoryear{{Kurokawa} \& {Kawai}}{{Kurokawa} \&
  {Kawai}}{1993}]{kurokawa93}
{Kurokawa}, H.,  \& {Kawai}, G. 1993, in Astronomical Society of the Pacific
  Conference Series, Vol.~46, IAU Colloq. 141: The Magnetic and Velocity Fields
  of Solar Active Regions, ed. {H.~Zirin, G.~Ai, \& H.~Wang}, 507

\bibitem[\protect\citeauthoryear{{Kurokawa} et~al.}{{Kurokawa}
  et~al.}{2007}]{kuro2007}
{Kurokawa}, H., {Liu}, Y., {Sano}, S.,  \& {Ishii}, T.~T. 2007, in Astronomical
  Society of the Pacific Conference Series, Vol. 369, New Solar Physics with
  Solar-B Mission, ed. {K.~Shibata, S.~Nagata, \& T.~Sakurai}, 347

\bibitem[\protect\citeauthoryear{{Lin} et~al.}{{Lin} et~al.}{2002}]{Lin02}
{Lin}, R.~P., et~al. 2002, \solphys, 210, 3

\bibitem[\protect\citeauthoryear{{Liu} \& {Kurokawa}}{{Liu} \&
  {Kurokawa}}{2004}]{Liu04}
{Liu}, Y.,  \& {Kurokawa}, H. 2004, \apj, 610, 1136

\bibitem[\protect\citeauthoryear{{Madjarska}, {Doyle}, \& {de
  Pontieu}}{{Madjarska} et~al.}{2009}]{Maj09}
{Madjarska}, M.~S., {Doyle}, J.~G.,  \& {de Pontieu}, B. 2009, \apj, 701, 253

\bibitem[\protect\citeauthoryear{{McIntosh} \& {De Pontieu}}{{McIntosh} \& {De
  Pontieu}}{2009}]{mcintosh09}
{McIntosh}, S.~W.,  \& {De Pontieu}, B. 2009, \apjl, 706, L80

\bibitem[\protect\citeauthoryear{{Moreno-Insertis}, {Galsgaard}, \&
  {Ugarte-Urra}}{{Moreno-Insertis} et~al.}{2008}]{Moreno08}
{Moreno-Insertis}, F., {Galsgaard}, K.,  \& {Ugarte-Urra}, I. 2008, \apj, 673,
  L211

\bibitem[\protect\citeauthoryear{{Murawski}, {Srivastava}, \&
  {Zaqarashvili}}{{Murawski} et~al.}{2011}]{Mur11}
{Murawski}, K., {Srivastava}, A.~K.,  \& {Zaqarashvili}, T.~V. 2011, \aap, 535,
  A58

\bibitem[\protect\citeauthoryear{{Nitta} \& {De Rosa}}{{Nitta} \& {De
  Rosa}}{2008}]{nitta08}
{Nitta}, N.~V.,  \& {De Rosa}, M.~L. 2008, \apj, 673, L207

\bibitem[\protect\citeauthoryear{{Pariat}, {Antiochos}, \& {DeVore}}{{Pariat}
  et~al.}{2010}]{pariat10}
{Pariat}, E., {Antiochos}, S.~K.,  \& {DeVore}, C.~R. 2010, \apj, 714, 1762

\bibitem[\protect\citeauthoryear{{Rachmeler} et~al.}{{Rachmeler}
  et~al.}{2010}]{rachmeler10}
{Rachmeler}, L.~A., {Pariat}, E., {DeForest}, C.~E., {Antiochos}, S.,  \&
  {T{\"o}r{\"o}k}, T. 2010, \apj, 715, 1556

\bibitem[\protect\citeauthoryear{{Roy}}{{Roy}}{1973}]{roy1973}
{Roy}, J.~R. 1973, \solphys, 28, 95

\bibitem[\protect\citeauthoryear{{Rust}}{{Rust}}{1968}]{rust1968}
{Rust}, D.~M. 1968, in IAU Symposium, Vol.~35, Structure and Development of
  Solar Active Regions, ed. {K.~O.~Kiepenheuer}, 77

\bibitem[\protect\citeauthoryear{{Sainz Dalda} \& {L{\'o}pez Ariste}}{{Sainz
  Dalda} \& {L{\'o}pez Ariste}}{2007}]{Dalda07}
{Sainz Dalda}, A.,  \& {L{\'o}pez Ariste}, A. 2007, \aap, 469, 721

\bibitem[\protect\citeauthoryear{{Scherrer} et~al.}{{Scherrer}
  et~al.}{1995}]{sche1995}
{Scherrer}, P.~H., et~al. 1995, \solphys, 162, 129

\bibitem[\protect\citeauthoryear{{Schmieder}, {Golub}, \&
  {Antiochos}}{{Schmieder} et~al.}{1994}]{Sch94}
{Schmieder}, B., {Golub}, L.,  \& {Antiochos}, S.~K. 1994, \apj, 425, 326

\bibitem[\protect\citeauthoryear{{Schmieder} et~al.}{{Schmieder}
  et~al.}{1984}]{schmieder84}
{Schmieder}, B., {Mein}, P., {Martres}, M.~J.,  \& {Tandberg-Hanssen}, E. 1984,
  \solphys, 94, 133

\bibitem[\protect\citeauthoryear{{Schmieder} et~al.}{{Schmieder}
  et~al.}{1988}]{schmieder88}
{Schmieder}, B., {Mein}, P., {Simnett}, G.~M.,  \& {Tandberg-Hanssen}, E. 1988,
  \aap, 201, 327

\bibitem[\protect\citeauthoryear{{Schmieder} et~al.}{{Schmieder}
  et~al.}{1995}]{schmieder95}
{Schmieder}, B., {Shibata}, K., {van Driel-Gesztelyi}, L.,  \& {Freeland}, S.
  1995, \solphys, 156, 245

\bibitem[\protect\citeauthoryear{{Shibata} et~al.}{{Shibata}
  et~al.}{1992}]{Shib92}
{Shibata}, K., et~al. 1992, \pasj, 44, L173

\bibitem[\protect\citeauthoryear{{Shibata} et~al.}{{Shibata}
  et~al.}{1982}]{Shib82}
{Shibata}, K., {Nishikawa}, T., {Kitai}, R.,  \& {Suematsu}, Y. 1982, \solphys,
  77, 121

\bibitem[\protect\citeauthoryear{{Shibata} et~al.}{{Shibata}
  et~al.}{1994}]{shibata1994}
{Shibata}, K., {Nitta}, N., {Strong}, K.~T., {Matsumoto}, R., {Yokoyama}, T.,
  {Hirayama}, T., {Hudson}, H.,  \& {Ogawara}, Y. 1994, APJL, 431, L51

\bibitem[\protect\citeauthoryear{{Srivastava} \& {Murawski}}{{Srivastava} \&
  {Murawski}}{2011}]{Sri11}
{Srivastava}, A.~K.,  \& {Murawski}, K. 2011, \aap, 534, A62

\bibitem[\protect\citeauthoryear{{Srivastava} \& {Murawski}}{{Srivastava} \&
  {Murawski}}{2012}]{Sri12}
{Srivastava}, A.~K.,  \& {Murawski}, K. 2012, \apj, 744, 173

\bibitem[\protect\citeauthoryear{{Sterling}}{{Sterling}}{2000}]{Ster20}
{Sterling}, A.~C. 2000, \solphys, 196, 79

\bibitem[\protect\citeauthoryear{{Sterling}, {Shibata}, \&
  {Mariska}}{{Sterling} et~al.}{1993}]{Ster93}
{Sterling}, A.~C., {Shibata}, K.,  \& {Mariska}, J.~T. 1993, \apj, 407, 778

\bibitem[\protect\citeauthoryear{{Uddin} et~al.}{{Uddin}
  et~al.}{2004}]{uddin04}
{Uddin}, W., {Jain}, R., {Yoshimura}, K., {Chandra}, R., {Sakao}, T., {Kosugi},
  T., {Joshi}, A.,  \& {Despande}, M.~R. 2004, \solphys, 225, 325

\bibitem[\protect\citeauthoryear{{Uddin} et~al.}{{Uddin}
  et~al.}{2010}]{Uddin10}
{Uddin}, W., {Kumar}, P., {Srivastava}, A.~K.,  \& {Chandra}, R. 2010, in
  Magnetic Coupling between the Interior and Atmosphere of the Sun, ed.
  {S.~S.~Hasan \& R.~J.~Rutten}, 478

\bibitem[\protect\citeauthoryear{{Ugarte-Urra} \& {Warren}}{{Ugarte-Urra} \&
  {Warren}}{2011}]{ugarte11}
{Ugarte-Urra}, I.,  \& {Warren}, H.~P. 2011, \apj, 730, 37

\bibitem[\protect\citeauthoryear{{Warren} et~al.}{{Warren}
  et~al.}{2011}]{warren11}
{Warren}, H.~P., {Ugarte-Urra}, I., {Young}, P.~R.,  \& {Stenborg}, G. 2011,
  \apj, 727, 58

\bibitem[\protect\citeauthoryear{{Yokoyama} \& {Shibata}}{{Yokoyama} \&
  {Shibata}}{1996}]{yokoyama1996}
{Yokoyama}, T.,  \& {Shibata}, K. 1996, Astrophysical Letters Communications,
  34, 133

\bibitem[\protect\citeauthoryear{{Yoshimura} et~al.}{{Yoshimura}
  et~al.}{2003}]{Yushi03}
{Yoshimura}, K., {Kurokawa}, H., {Shimojo}, M.,  \& {Shine}, R. 2003, \pasj,
  55, 313

\end{thebibliography}
\clearpage

\begin{figure}
\centering{
\includegraphics[width=12cm]{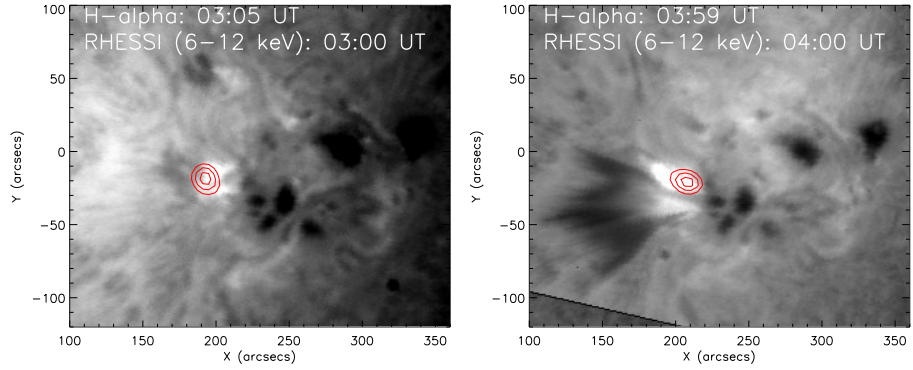}
} 
\caption{H$\alpha$ images overlaid by RHESSI 6-12 keV contours
during Flare 2 and Flare 3 (bump 3)
 The contour levels are 50,
70 and 90$\%$ of the peak intensity.} \label{fig5}
\label{fig4}
\end{figure}
\begin{figure}
\centering{
\includegraphics[width=12cm,clip=]{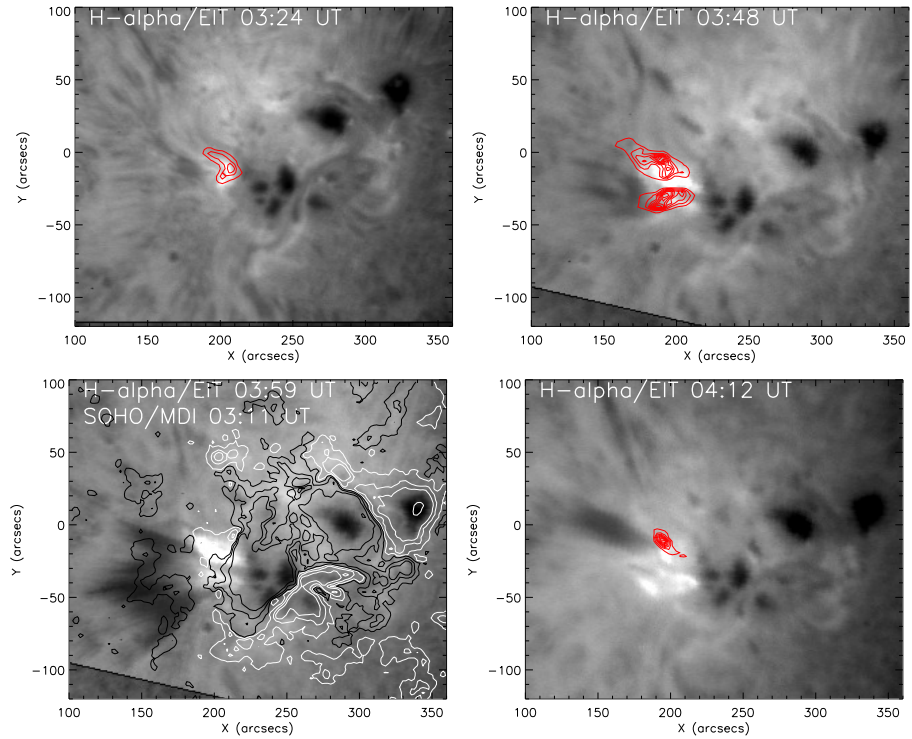}
}

\caption{H$\alpha$ images showing the flares and surge
eruptions  overlaid by the SOHO/EIT 171 \AA\ contours  for three
different times (top, and bottom right panel). The bottom left image
shows an H$\alpha$ image overlaid by MDI contours (white contours
show the positive polarity whereas black contours indicate the
negative polarity).}
\end{figure}
\begin{figure}
\centering{
\hspace*{-0.06\textwidth}
\includegraphics[width=10cm]{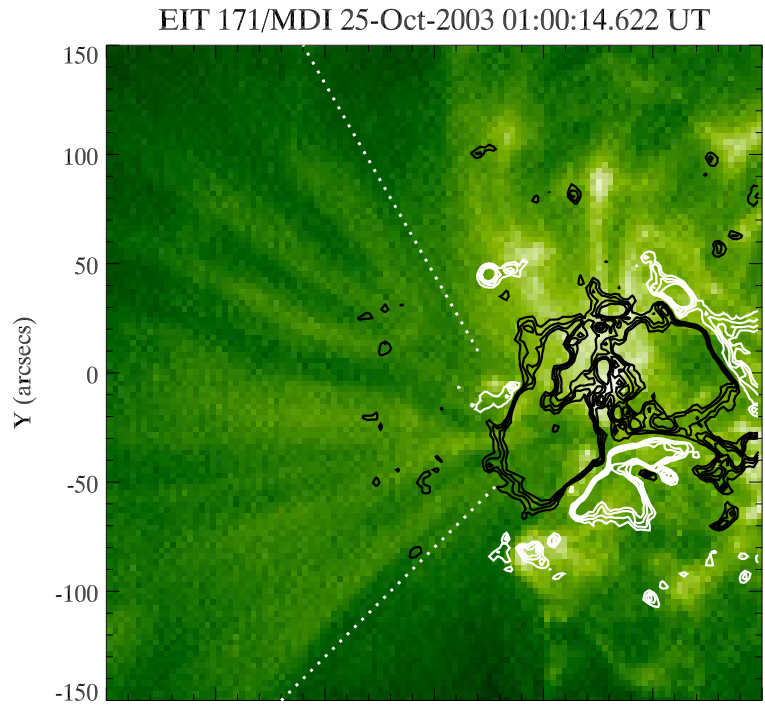}
}
\caption{SOHO/EIT 171 \AA \ image overlaid by the MDI contours
 (white shows the positive polarity whereas black indicates the negative
  polarity). The dotted lines indicate the fan-shaped structure  of  field lines over the activity
  site before the flare
and surge activities.}
\label{fig3}
\end{figure}

\begin{figure}[htbp]
\begin{center}
\includegraphics[width=12cm]{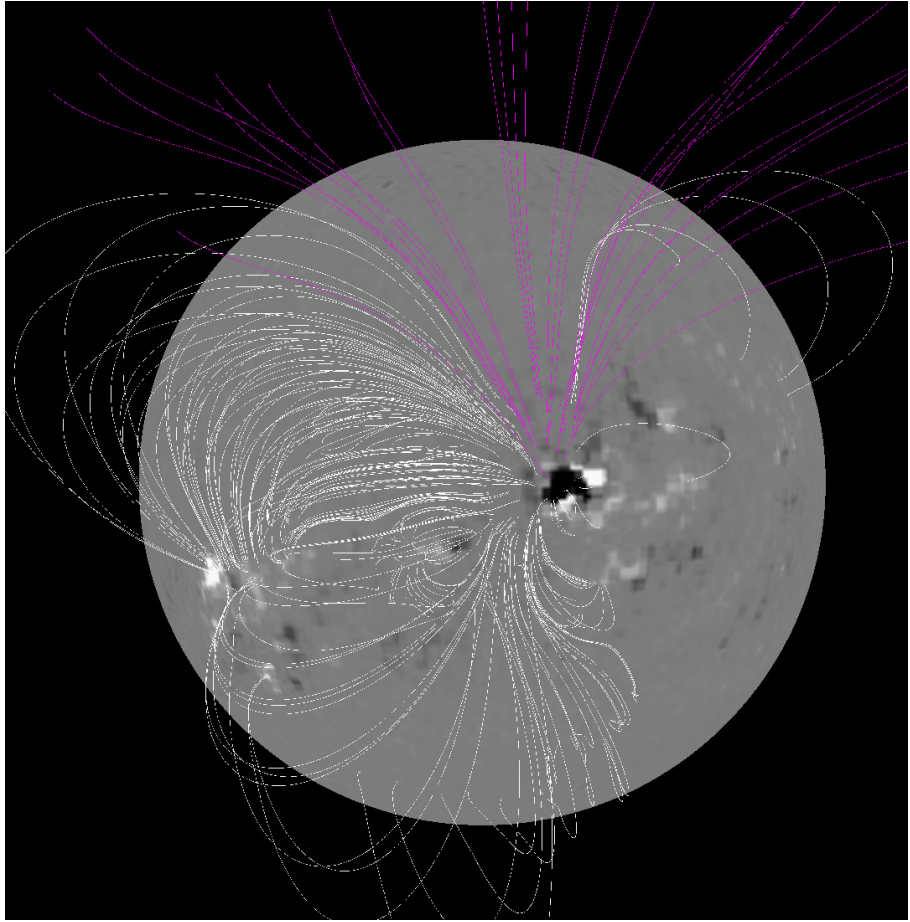}
\caption{Potential Field Source Surface (PFSS) extrapolation overplotted at the SoHO/MDI
full disk map on October 25 2003 on 00:04 UT, which shows
the large-scale field connectivity of two active regions as well as the
topology of magnetic field near the surge productivity site in AR10484.}
\label{fig4-bis}
\end{center}
\end{figure}

\begin{figure}[htbp]
\begin{center}
\includegraphics[width=12cm]{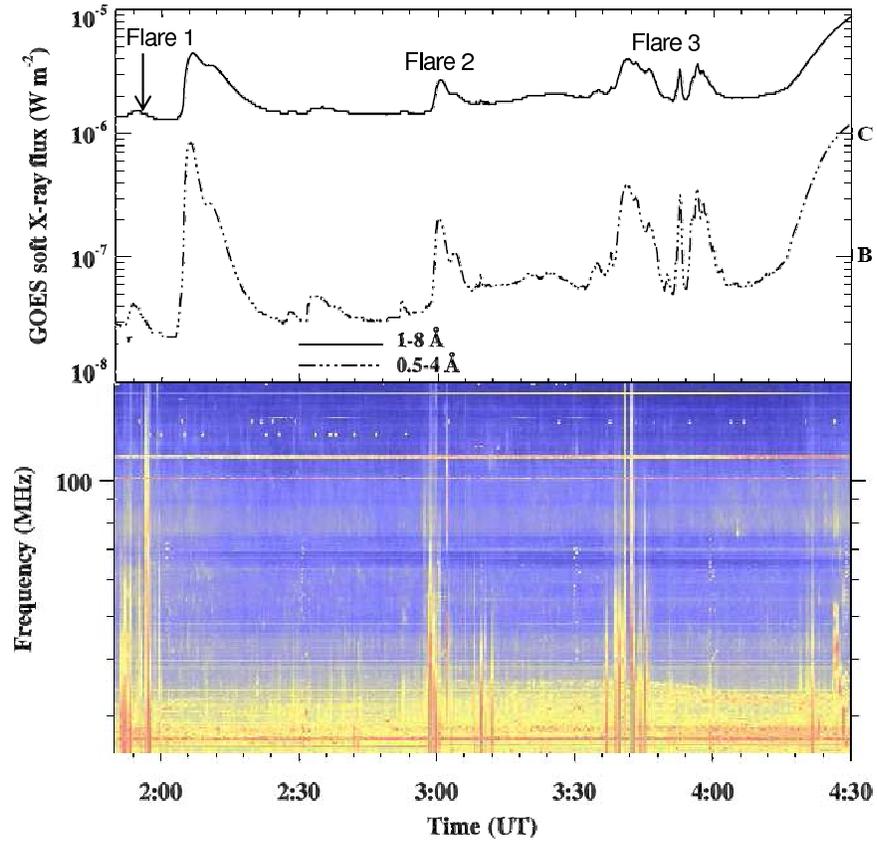}
\caption{GOES soft X-ray light  curve and Radio TYPE III bursts ( Learmonth)}
\label{fig4-ter}
\end{center}
\end{figure}


\begin{figure}
\centering{
\includegraphics[width=18cm]{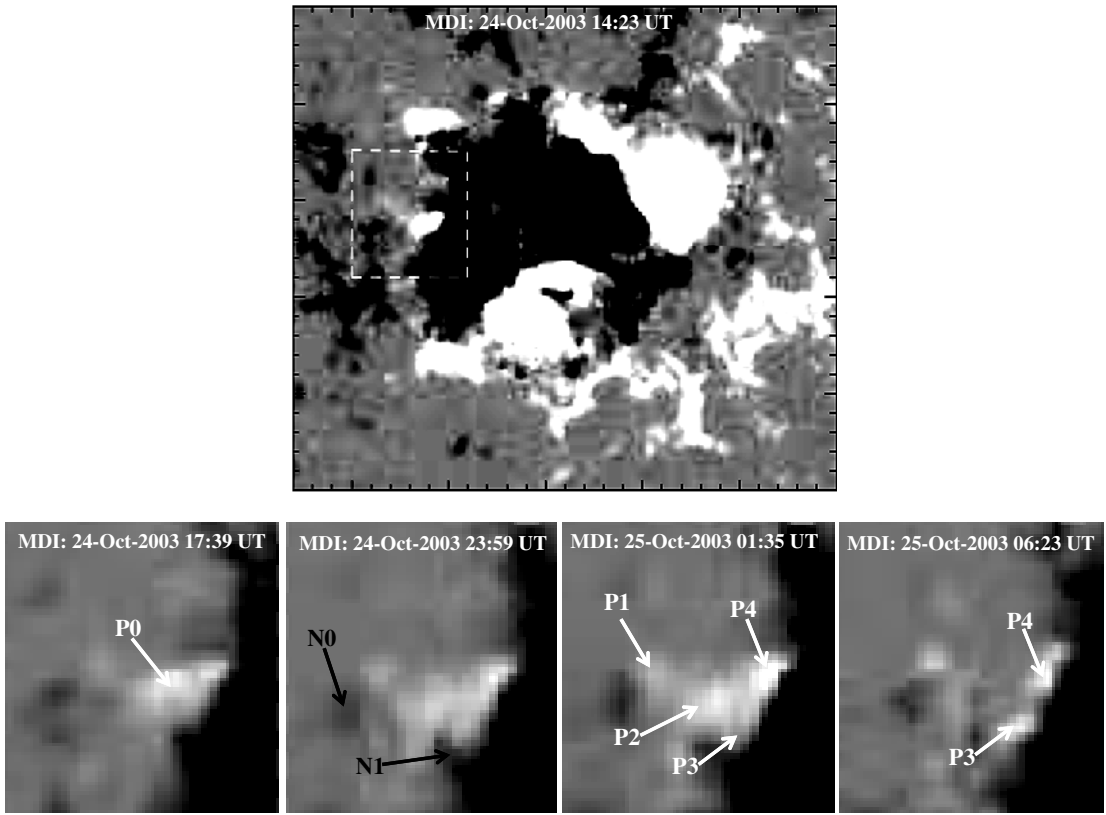}
} \vspace*{-1.5cm} \caption{Magnetogram on 24 October, 2003 before
the surge activity showing the moat region around the main
polarities (top panel) and zoom on the region of the surges
(presented as a white box in the top panel)  before, during, and
after the surges (bottom panel). The field-of-view of the upper and
lower images is 280$''\times 250''$ and $60''\times65''$
respectively.} \label{mdi_evo}
\end{figure}

\begin{figure}
\centering{
\includegraphics[width=15cm]{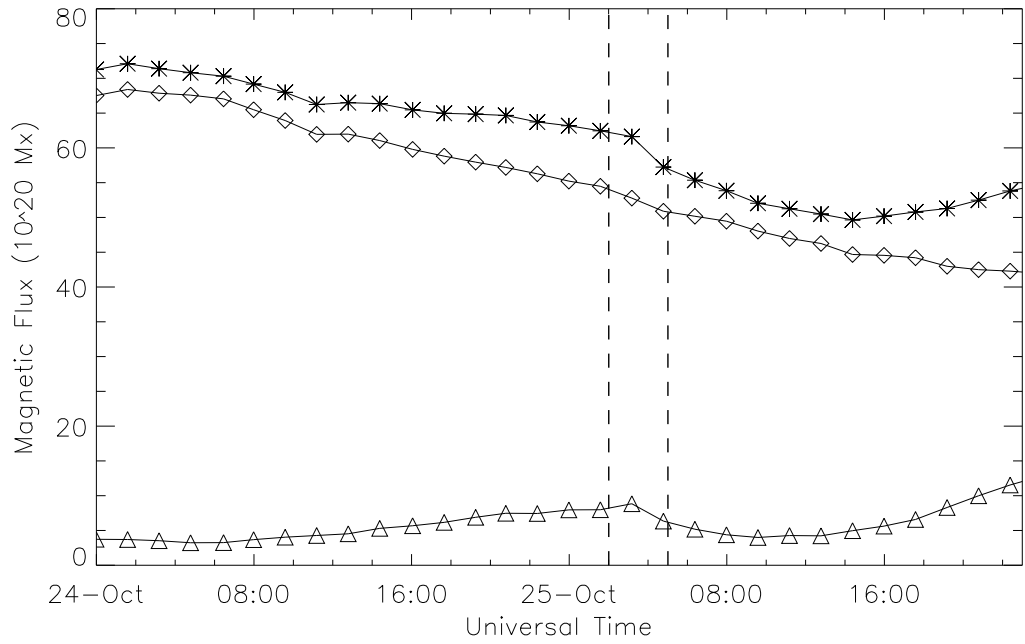}
} \caption{The temporal variation of positive (triangle), negative
(diamond) and total (star) magnetic flux before, during, and after the surge
activities as measured over the selected area shown in the bottom
panels of Figure \ref{mdi_evo}. The two vertical lines  indicate the
duration of surge activity.} \label{flux}
\end{figure}
\begin{figure}
\centering{
\hspace*{1cm}
\includegraphics[width=18cm]{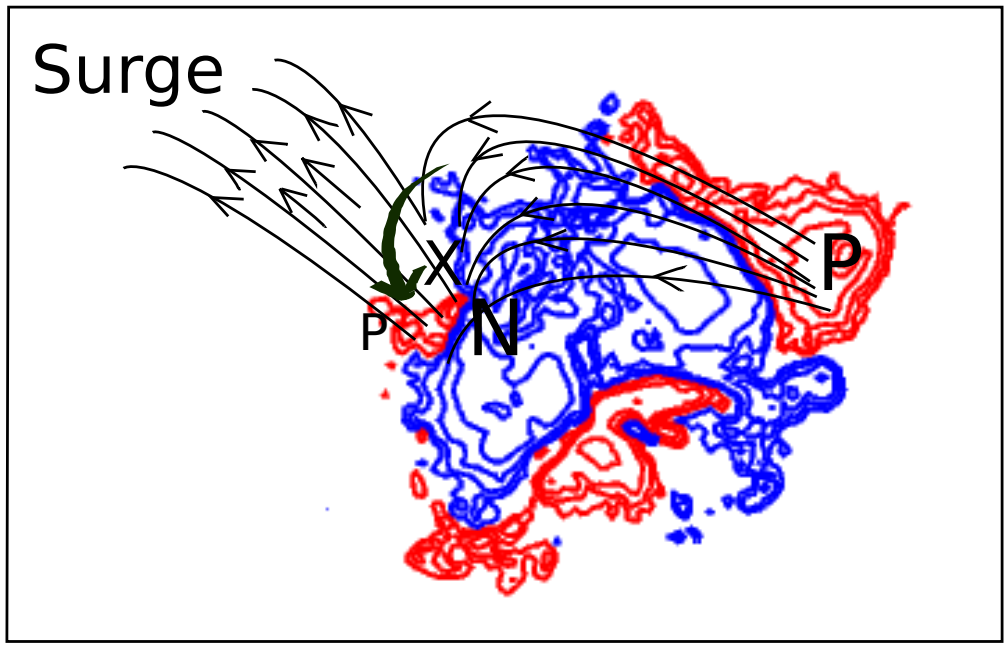}
} \vspace*{-1cm} \caption{Schematic cartoon showing the surge
initiation due to the reconnection (shown by $'$X$'$) of the field
lines with the opposite polarity field region, resulting of  the
anticlockwise motion of the negative polarity field region
(indicated by thick arrow). Red contours indicate the positive
polarity whereas blue ones show the negative polarity field region.
Surge moves along the open field lines.} \label{cartoon}
\end{figure}


\end{document}